\input epsf

\documentstyle[preprint,prd,aps,rotate]{revtex}

\begin{document}
\draft
\title{Detectability of gravitational wave events by 
       spherical resonant-mass antennas}
\author{Gregory M. Harry\cite{gemail}, Thomas R. 
        Stevenson\cite{temail}, 
        Ho Jung Paik\cite{hemail}}
\address{Department of Physics, University of Maryland, 
         College Park, MD
         20742}
\date{\today}
\maketitle

\begin{abstract}
We report on results of computer simulations of spherical 
resonant--mass gravitational wave antennas interacting with 
high--frequency 
radiation from astronomical sources.  The antennas were simulated 
with three--mode inductive transducers placed on the faces of
a truncated icosahedron.  Overall, the spheres were modeled with a 
sensitivity of about three times the standard quantum limit.  The 
gravitational radiation data used was generated 
by  three--dimensional numerical computer models of inspiraling 
and coalescing binary neutron stars and of the dynamical bar--mode 
instability of a rapidly rotating
star.  We calculated energy 
signal--to--noise ratios
for aluminum spheres of different sizes cooled to 50~mK.  
We find that by using technology 
that could be available in the next several years, spherical antennas 
can detect coalescing binaries out to slightly over 15 Mpc, the lower 
limit on the distance required for one event per
year.  For the rapidly rotating star, we find, for a particular
choice of the radius at centrifugal hangup, spheres are sensitive 
out to about 2 Mpc.  The event rate is estimated to be about 1 
every 10 years at this distance.
\end{abstract}
\section{Introduction}

The experimental effort to directly detect the space--time ripples 
known as 
gravitational waves has been going on for 35 years, beginning with 
Weber's
pioneering work in the early 1960s \cite{Weber}.  Since then, two 
main experimental
approaches have evolved: cryogenic resonant--mass detectors 
\cite{LSU,CERN,Aussies} and
laser interferometers \cite{LIGO,VIRGO,GEO}.  The sensitivity of 
both techniques 
is sufficient that unambiguous detection of gravitational waves 
is expected soon, perhaps 
within the next ten years.  The work of Taylor and Hulse 
\cite{Taylor}, showing the orbital 
decay of binary pulsar PSR~1913-16 is in agreement with general 
relativity's prediction for 
gravitational wave emission, has added to the
anticipation of the first direct, confirmed detection. It appears 
possible that a new generation of advanced resonant--mass detectors 
could operate concurrently with interferometers already under
construction.  At this time, understanding possible
sources of gravitational waves and which experimental technique 
is best suited to which
source of radiation takes on greater importance. 
We endeavor to clarify this by numerically computing 
energy signal--to--noise ratios for resonant--mass detectors and 
interferometers interacting with two possible sources of 
gravitational waves.

The best understood source of detectable gravitational waves is  
from inspiraling and coalescing binary neutron stars \cite{Phinney}.
The Laser Interferometer Gravitational--wave Observatory (LIGO) has 
been designed and 
optimized to detect these events at a distance of 200 Mpc after 
significant interferometer improvements \cite{LIGO}. To accomplish this, 
it has been devised 
to be most sensitive at as low a frequency as possible 
($\sim 200$ Hz) where the waveform
from binary neutron stars is stronger. However, the waveform at 
200 Hz is due 
almost solely to the {\em inspiral} phase of 
the binary neutron star evolution and contains virtually no 
information about the
{\em coalescence}.  As the inspiral is determined by point-mass dynamics, 
the equation
of state for nuclear matter (i.e., neutron stars) will affect only 
the
coalescence waveform.  Coalescence also occurs when the gravitational 
field
between the neutron stars is strongest, so the effects of general 
relativity will
be more important than during the inspiral.  To measure these effects, 
it will be
necessary to monitor the {\em higher--frequency}
waves from coalescence in addition to those at lower frequency from 
inspiral.

Resonant-mass gravitational wave detectors have been in use 
for longer than the interferometric detectors. 
Resonant--mass antennas with bar
geometries have been taking data and been continually improved since 
their inception.
The use of spherical geometry as an improvement over bars was first 
suggested by 
Forward in 1971 \cite{Forward}, and Wagoner and Paik later showed that 
at equal frequencies 
spheres have an advantage over bars in energy cross section 
\cite{Wagoner}.  
Recent re-examination of spherical detectors \cite{JM,ZM,MJ,Coccia}
has generated international interest in constructing 
one or more massive spherical antennas incorporating advances in 
transducer technology.  The possibility of building such an antenna 
to operate concurrently with the first LIGO 
interferometers appears good \cite{coop}.  
The sensitive frequencies for a sphere are higher than 
those for the first LIGO interferometers, spanning about 750~Hz 
to 2700~Hz in the lowest mode,
and therefore they are well suited
to complement interferometer experiments at high frequencies.  
One such advanced resonant--mass detector can be more 
sensitive than the first LIGO interferometers within a bandwidth of 
around 100~Hz centered 
at the quadrupole 
resonance of the sphere and would have a sensitivity within that
band comparable to what will be achieved by the advanced LIGO 
interferometers. The sphere's sensitivity is independent of
source direction, unlike LIGO.   Spherical antennas can also provide 
direction and polarization information 
more easily than LIGO \cite{ZM} and, properly instrumented, could 
detect any scalar gravitational radiation that might be present 
\cite{Forward,Wagoner}.  In this paper, when we refer to 
``high--frequency''
gravitational radiation, we mean those signals that include 
significant strength above
750~Hz. This frequency is where the first LIGO interferometers' 
sensitivity begins to weaken from photon shot 
noise in the lasers \cite{LIGO} and the spherical resonant--mass 
detector's sensitivity becomes important.

We have looked at the question of whether a spherical detector, or in 
particular a Truncated
Icosahedral Gravitational-wave Antenna (TIGA)  as
described by Merkowitz and Johnson \cite{MJ}, is capable of 
observing high--frequency events. Specifically, the coalescence of a 
binary neutron
star system and the dynamical bar--mode instability of a single, 
rapidly rotating star were examined
as possible astronomical phenomenon that could produce 
high--frequency gravitational radiation.  
Waveforms for these events, generated with computer simulations by 
Centrella's
group at Drexel University, were used as input into a mathematical 
model of a 50~mK 
spherical detector with  three--mode inductive transducers 
\cite{Paik}.  The energy
signal-to-noise ratios obtained from this model help determine how 
TIGAs 
and interferometer experiments can best complement one another.

Coccia and Fafone \cite{Fafone} have also looked at energy
signal--to--noise ratios 
from 
astronomical events in spherical detectors.  Our work and theirs 
are complementary.  They looked solely at inspiraling binary 
neutron stars as sources, 
leaving out the coalescence phase as well as any other 
high--frequency events.  Since 
the inspiral can be modeled accurately by point--mass dynamics, they 
used an 
analytical 
expression for the waveform.  We found it necessary to use numerical 
data from 
computer models 
to simulate the coalescence.  By limiting themselves to the inspiral 
phase, Coccia and Fafone were unable to accurately predict energy
signal--to--noise 
ratios for higher mass 
neutron stars or black holes. For some sphere sizes and compositions 
their 
simulation 
does not produce results for 1.4 M$_{\odot}$ neutron stars, the 
observed 
mass for all 
known neutron stars in binaries \cite{Phinney}.  However, their 
method was 
able to show that spherical 
antennas can determine the chirp mass 
($M_{c} \equiv (M_{1} M_{2})^{3/5}/(M_{1} + M_{2})^{1/5}$
\cite{Thorne}), the orbital inclination and the distance to the 
source, a result our method did not produce.  Both techniques provide 
useful information that are unobtainable by the other.

In Section~\ref{method}, we describe the method used for the energy
signal-to-noise ratio calculations,
how the code that produced the results was written, and what parameters 
for the spherical antenna we used. 
In Section~\ref{results},  we discuss the signal waveforms we used as 
inputs into the model 
developed in Section~\ref{method} and present the results of the 
calculations.  
Finally, in Section~\ref{con}, we present our conclusions and discuss 
ideas for 
further work.

\section{Method}
\label{method}

To calculate the energy signal--to--noise ratio (SNR) per unit 
bandwidth of 
the TIGA, we followed the 
method of Price \cite{Price} who showed how to calculate the SNR for a 
bar 
antenna that uses
an optimal filter to process the data.  Stevenson \cite{Thomas} has 
shown that for six identical 
transducers in the TIGA  geometry and identical Qs for all five 
quadrupole modes of the sphere, 
the SNR of a spherical antenna is identical to that of a bar antenna 
instrumented with one of those six transducers.   
The {\em equivalent bar} has an effective mass as seen by the 
transducer of

\begin{equation}
m_{\text{eff}} = \frac{5}{6} \chi (\frac{4}{3} \pi R^{3}) \rho,
\label{meff}
\end{equation}
where $R$ and $\rho$ are the sphere radius 
and density.  The dimensionless parameter $\chi$ comes from the
radial 
driving point admittance matrix of the sphere at the quadrupole 
frequency.  For an aluminum sphere with a Poisson ratio of~0.33, 
$\chi = 0.301$ \cite{ZM}. The factor
of 5/6 in Eq.~(\ref{meff}) accounts for the multiple sphere modes and 
transducers 
\cite{Thomas} .    
The SNR for the TIGA is the same as that for the equivalent bar 
provided one equates the energies deposited by the gravitational wave 
in the two antennas.  We can now calculate the SNR for 
the simpler case of a bar,  while retaining all the information 
available from a sphere.

The energy deposited in the sphere is calculated from
\begin{equation}
E = {\mathcal F}_{\omega} \Sigma,
\end{equation}
where $\Sigma$ is the energy cross section of the sphere \cite{ZM},
\begin{equation}
\Sigma = \frac{G}{c^{3}} \frac{\rho V_{s}^{5}}{f_{0}^{3}} \Pi.
\label{cross}
\end{equation}
Here  $V_{s}$ is the extensional sound speed of the sphere material, 
$f_{0}$ is the 
quadrupole frequency, and $\Pi$ is a dimensionless constant that 
accounts for antenna geometry and
mode.  It has the value 0.215 \cite{ZM} for a sphere in the lowest 
quadrupole mode and 0.585 in the
first excited quadrupole mode \cite{ZM,Coccia}.  Throughout $G$ and $c$ 
are Newton's gravitational constant and the speed of light. The 
total energy flux 
$\cal{F}_{\omega}$ is \cite{MTW}

\begin{equation}
{\mathcal F}_{\omega} = \frac{c^{3}}{G} \frac{1}{16 \pi} 
                        \omega^{2} |h(\omega)|^{2},
\end{equation}
where $\omega$ is the angular frequency of the gravitational radiation 
and $|h(\omega)|$ is
the magnitude of the frequency--domain amplitude of the gravitational 
wave.
Thus the total energy deposited in the equivalent bar is 

\begin{equation}
E = \frac{\pi}{2} \frac{\rho V_{s}^{5}}{f_{0}} 
         \Pi |h(\omega_{0})|^{2}.
\label{energy}
\end{equation}
We define an {\em effective force} that acts on the equivalent 
bar \cite{ZM}:

\begin{equation}
f_{\text{eff}}(\omega) = - m_{\text{eff}} \, \omega^{2} h(\omega) \ell_{\text{eff}},
\label{feff}
\end{equation}
where the relationship between $\ell_{\text{eff}}$ and $R$ is determined as follows.
For an impulsive force, the energy is deposited as kinetic energy in the 
antenna.  The energy $E$ after the impulse is given by
\begin{equation}
E= \frac{|f_{\text{eff}}(\omega_{0})|^{2}}{2 m_{\text{eff}}}.
\label{energy2}
\end{equation}
Combining Eqs.~(\ref{energy}-\ref{energy2}) gives 
\begin{equation}
\ell_{\text{eff}}  =  \sqrt{\Pi \frac{2 \pi^{2} \rho V_{s}^{5}}
                                       {m_{\text{eff}} \omega_{0}^{5}}}.
\end{equation}
Then using Eq.~(\ref{meff}) and the relationship between $\omega_{0}, V_{s},$ 
and $R$ \cite{Love}
for each quadrupole mode an $\ell_{\text{eff}}$ of $0.337 R$ in the lowest 
quadrupole mode and $0.109 R$ in the first excited quadrupole mode are 
calculated.  Using Eq.~(\ref{feff}) as the definition of a force on the 
equivalent bar, the method of Price can be followed exactly.

The transducer we assumed was a three--mode inductive 
transducer.  A three--mode 
transducer is necessary, rather than the standard two--mode system, to 
get higher bandwidths,
which are required to reach sensitivities near the standard quantum 
limit.  Higher bandwidth
reduces the requirement on the Q of the sphere and transducer.  Higher 
bandwidth is also useful to cover a larger spectrum 
of frequencies and reduce the 
need for additional antennas. 

Assuming that a template of the gravitational waveform is available, 
optimal filtering 
can be used on the output signal of the transducer. The optimal 
filter produces the highest 
SNR possible \cite{Wainstein} and has the form

\begin{equation}
K(\omega) = \frac{e^{-j \omega t_{0}} u^{*}(\omega)}{S_{n}(\omega)},
\end{equation}
where $u(\omega)$ is the velocity signal of the antenna effective mass 
and 
$S_{n}(\omega)$ is the 
total velocity noise spectral density, both referred to the input of 
the optimal filter.  The parameter
$t_{0}$ is the time at which the SNR is to be optimized.  To calculate 
$u(\omega)$ and $S_{n}(\omega)$, it is necessary to solve the equations 
of motion for the 
antenna coupled to the three--mode resonant transducer.  They have the 
form:

\begin{eqnarray}
j \omega m_{\text{eff}} u_{1}& = 
               &f_{1}- \frac{j k_{\text{int}}}{\omega} (u_{\text{int}}
               - u_{1}) + \frac{j k_{\text{eff}}}{\omega} u_{1}, 
                 \nonumber \\
j \omega m_{\text{int}} u_{\text{int}} & = 
               &  \frac{j k_{\text{int}}}{\omega} (u_{\text{int}}
                - u_{1})-\frac{j k_{\text{trans}}}{\omega} 
                (u_{2}-u_{\text{int}}) - f_{2},  \label{eqnmotion} \\ 
j \omega m_{\text{trans}} u_{2}        & = 
               & \frac{j k_{\text{trans}}}{\omega} (u_{2} - 
                u_{\text{int}}) + f_{2}. \nonumber
\end{eqnarray}
Here, $m_{\text{eff}}$ is the effective mass of the antenna, 
$m_{\text{int}}$ 
is the mass of the intermediate 
resonator, and $m_{\text{trans}}$ is the transducer mass; $k_{\text{eff}}$ 
is the 
effective spring constant of 
the antenna, $k_{\text{int}}$ is the spring constant that connects the 
antenna to the intermediate mass, 
and $k_{\text{trans}}$ is the spring constant between the intermediate 
mass and 
the transducer mass.  
The spring constants are complex valued and include dissipation.  The 
variables $u_{1}, u_{\text{int}}$, and $u_{2}$ are, respectively, the 
velocities 
of the antenna 
surface at the transducer,  of the intermediate mass, 
and of the transducer mass. Applied forces acting on the antenna 
surface and between
$m_{\text{trans}}$ and $m_{\text{int}}$ are denoted by $f_{1}$ and 
$f_{2}$, 
respectively.

Eliminating $u_{\text{int}}$ from Eq. (\ref{eqnmotion}) allows the 
equations of 
motion to be written
as
\begin{equation}
u_{i} = y_{ij} f_{j},
\label{redeqnmotion}
\end{equation}
with $i$ and $j$ taking the values 1 to 2. The energy SNR per unit bandwidth, 
$\sigma(\omega)$, then becomes
\cite{Price}
\begin{eqnarray}
\sigma(\omega) & = & K(\omega) u_{2} \\
                       & = & \frac{|u_{2}(\omega)|^{2}}{S_{n}(\omega)}.
\end{eqnarray}
From Eq. (\ref{redeqnmotion}), $\sigma(\omega)$ is found to be 
\begin{equation}
\sigma(\omega) = \frac{|f_{1} y_{21}|^{2}}{S_{u} + S_{f} |y_{22}|^{2} +
                   2 k_{B} T \, \text{Re}(y_{22}) + 2 \text{Re}(y_{22} 
                   S_{fu})},
\label{sigma}
\end{equation}
assuming no force on the transducer, i.e., $f_{2} =0$. Here the force 
$f_{1}$ 
is $f_{\text{eff}}$
from Eq. (\ref{feff}) and $T$ is the physical temperature of the sphere.  The 
matrix $y_{ij}(\omega)$ 
is the admittance matrix of the antenna with transducer defined in
Eq.~(\ref{redeqnmotion}).   The 
four terms in the denominator are the individual parts of 
$S_{n}(\omega)$, the 
velocity noise.  They 
are, respectively, the additive velocity noise, the force noise, the 
thermal 
noise, and the correlation 
noise.  The spectral densities are defined as
\begin{eqnarray}
S_{f}(\omega) & \equiv & \int^{+ \infty}_{- \infty} e^{- j \omega \tau} 
                         \langle f(t) 
                         f(t - \tau) \rangle d\tau, \nonumber \\
S_{u}(\omega) & \equiv & \int^{+ \infty}_{- \infty} e^{-j \omega \tau} 
                         \langle u(t) 
                         u(t - \tau) \rangle d\tau, \\ 
S_{fu}(\omega) & \equiv & \int^{+ \infty}_{- \infty} e^{- j \omega \tau} 
                         \langle f(t) 
                         u(t - \tau) \rangle d\tau. \nonumber
\end{eqnarray}

In practice, these noise terms are found not to vary much with frequency 
in the antenna's sensitive range.  It is often convenient to parametrize
these spectral densities with three values; noise temperature $T_{n}$,
noise resistance $r_{n}$, and noise reactance $x_{n}$.  They are defined as

\begin{eqnarray}
T_{n} & = & \frac{1}{k_{B}} \sqrt{S_{f} S_{u} - 
                                  [\text{Im}(S_{fu})]^{2}}, 
                   \label{tn} \\
r_{n} & = & \sqrt{\frac{S_{f}}{S_{u}} - 
                   \left( \frac{\text{Im}(S_{fu})}{S_{u}} \right)^{2}},
                   \label{rn} \\
x_{n} & = &  \frac{\text{Im}(S_{fu})}{S_{u}},
                   \label{xn}
\end{eqnarray}
where $k_{B}$ is Boltzman's constant.  For simplicity, we set 
$\text{Im}(S_{fu})~=~0$.  Although, in general,
the correlation between the force and velocity noise is non--zero, the 
effects of a non--zero 
$S_{fu}$ can normally be accounted for by a renormalization of the 
transducer spring constant \cite{Price}.  The real part of $S_{fu}$
is normally zero when a SQUID amplifier is used. 

Once a complete expression for $\sigma(\omega)$ has been obtained, the 
energy SNR can be calculated from
\begin{equation}
\text{SNR} = \frac{1}{2 \pi}\int^{+ \infty}_{- \infty} \sigma(\omega) d\omega.
\label{snr}
\end{equation}
Note that we have consistently used a double--sided spectral density 
in contrast to
the single--sided convention adopted by LIGO.  By putting in numerical 
values for all parameters, this integral can 
be evaluated.  For many of the parameters below, we chose values beyond
what has been demonstrated experimentally so as to represent an 
advanced spherical detector.  Such an advanced detector could operate 
concurrently with the 
first LIGO interferometer as a result of aggressive research and 
development efforts now being planned \cite{coop}.  Some of our parameter
values are only slight extrapolations beyond currently demonstrated values,
while others are instead upper bounds to the technologies being pursued.
A detailed consideration of the research burden to meet each of our
assumed values is beyond the scope of this paper.  As stated in the 
introduction, our motivation is to
clarify how such spherical detectors could complement the interferometer
experiments by examining the detectability of high frequency events.
We feel it is
likely that resonant--mass detectors with an energy sensitivity approaching
that derived with our parameters can be developed and built to operate 
within the time frame between the completion of the first LIGO interferometers
and the operation of advanced interferometers.  

We chose to model aluminum 
spheres at a physical temperature of 50~mK, instrumented with six 
identical 
sets of three--mode
inductive transducer systems located with the dodecahedral TIGA 
geometry 
\cite{MJ}.  The lowest temperature that an aluminum
bar antenna has been successfully cooled to is 95~mK \cite{NAUTILUS}.  
Two--mode transducers are in
use on a number of operating cylindrical resonant--mass antennas 
\cite{LSU,CERN,Aussies} and a 
three--mode system has been demonstrated at 4 K in a smaller, test 
antenna \cite{Folkner}. 
A constant mass ratio between the effective mass of 
the sphere and the intermediate mass as well as between the 
intermediate and transducer masses 
of 100:1 was used, and all mechanical quality factors (Qs) were 
assumed to be
$40 \times 10^{6}$.  The highest mechanical Q that has been obtained 
in an inductive transducer is $24 \times 10^{6}$ \cite{Geng}.

The transducer electronics 
were assumed to be a 9 cm diameter inductive pickup coil attached to 
a SQUID amplifier with a quantum--limited noise temperature, 
i.e., $T_{s} = 1 \, \hbar \omega_{0}/k_{B}$.  
Quantum--limited SQUIDs have been
constructed \cite{Awschalom}, but they are not useful for 
inductive 
transducers because of their low input coil inductance.  Getting a 
suitable quantum--limited SQUID
is an area of intense research.  Wellstood's group at the 
University of Maryland is developing a quantum--limited 
SQUID for use in a gravitational wave
transducer.  The best noise temperature they have achieved in a SQUID 
with high
enough inductance to couple to the transducer coils is 
$T_{s} \approx 28 \, \hbar \omega_{0}/k_{B}$ 
\cite{Wellstood}.  The prospect of approaching the quantum--limit in a
practical SQUID in the next several years looks real.  With proper
matching,  the transducer noise is limited by
the noise of the SQUID, so the value of $T_{n}$ in 
Eq.~(\ref{tn}) becomes equal to $T_{s}$.  
The noise resistance is
$r_{n} = k_{E} /4 \pi f_{0}$,  where $k_{E}$ is the real part of the 
spring constant $k_{\text{trans}}$
that is due to the electrical interaction between the transducer mass 
and the pickup coils.  The
ratio $k_{E}/k_{\text{trans}}$ is the coupling between the electrical and 
mechanical parts of the
transducer.  For the value $k_{E}$, we took the product 
$3.78~\times~10^{8}$~N/m$^{3}~\times$~coil area, 
based on measurements made in our laboratory at Maryland
\cite{math}.

Taken together, these parameters define the overall sensitivity 
of the antenna.  Energy sensitivity $E_{s}$ is defined as
\begin{equation}
E_{s} = \frac{E}{\text{SNR}}.
\label{es}
\end{equation}
It is useful to
express this sensitivity in relation to the standard quantum limit, the
minimum sensitivity possible using a linear amplifier \cite{Giffard}.  
Expressed as a multiple of this
standard quantum limit, the antenna sensitivity becomes
\begin{equation}
N  =  \frac{E_{s}}{\hbar \omega_{0}}. 
\end{equation}
As a comparison for the numerical result, we calculate an approximate
value of $N$ from
\begin{equation}
 N  \approx  \frac{k_{B}}{\hbar \omega_{0}} \left[ T_{s} + 
              \frac{T}{\delta}  
              \left(\frac{1}{\alpha_{1} Q_{\text{eff}}} + 
              \frac{1}{\alpha_{2} Q_{\text{int}}} + 
              \frac{1}{\alpha_{3} Q_{\text{trans}}}  \right) \right].
\label{alpha}
\end{equation}
Equation (\ref{alpha}) is derived in the Appendix.  The parameters
$\delta$ and $\alpha_{i}$ are also defined and computed in the Appendix.
Substituting the values of $T, T_{s}, Q_{\text{eff}}, Q_{\text{int}},$ and 
$Q_{\text{trans}}$  assumed above
into Eq. (\ref{alpha}) gives
\begin{eqnarray}
 N & \approx & 1.0 + 0.96 + 0.87 + 0.78  \nonumber \\
   & \approx & 3.6. 
\end{eqnarray}

We calculated SNRs for eight different spheres.  The diameter of the 
lowest--frequency sphere 
was chosen to be the largest size that might be constructed, 3.25~m.  
The size of the highest--frequency sphere was chosen so that its 
lowest quadrupole frequency 
coincides with the peak in the spectrum of the coalescing binary 
neutron star data. 
This peak is at twice the rotation frequency of the 
transient, barlike structure that 
forms immediately after coalescence \cite{Zhuge}.  This assumption 
gives a sphere diameter of about 1.05~m.  The 
remaining sphere sizes were chosen to give reasonably continuous 
coverage of the frequency band 
750~Hz to 2700~Hz.  In addition to transducers tuned to the lowest 
quadrupole mode of the sphere, a
system tuned to the first excited mode was examined.  Coccia, Lobo, 
and Ortega 
\cite{Coccia} have shown that the cross section of the excited mode 
of a large sphere is 2.72 
times greater than that for the lowest mode of a small sphere at the 
same resonance frequency.   This
allows the calculations of SNRs for the higher mode.

\begin{figure}[p] 
\begin{center}
\leavevmode
\rotate[l]{\epsfxsize=3.5in\epsfbox{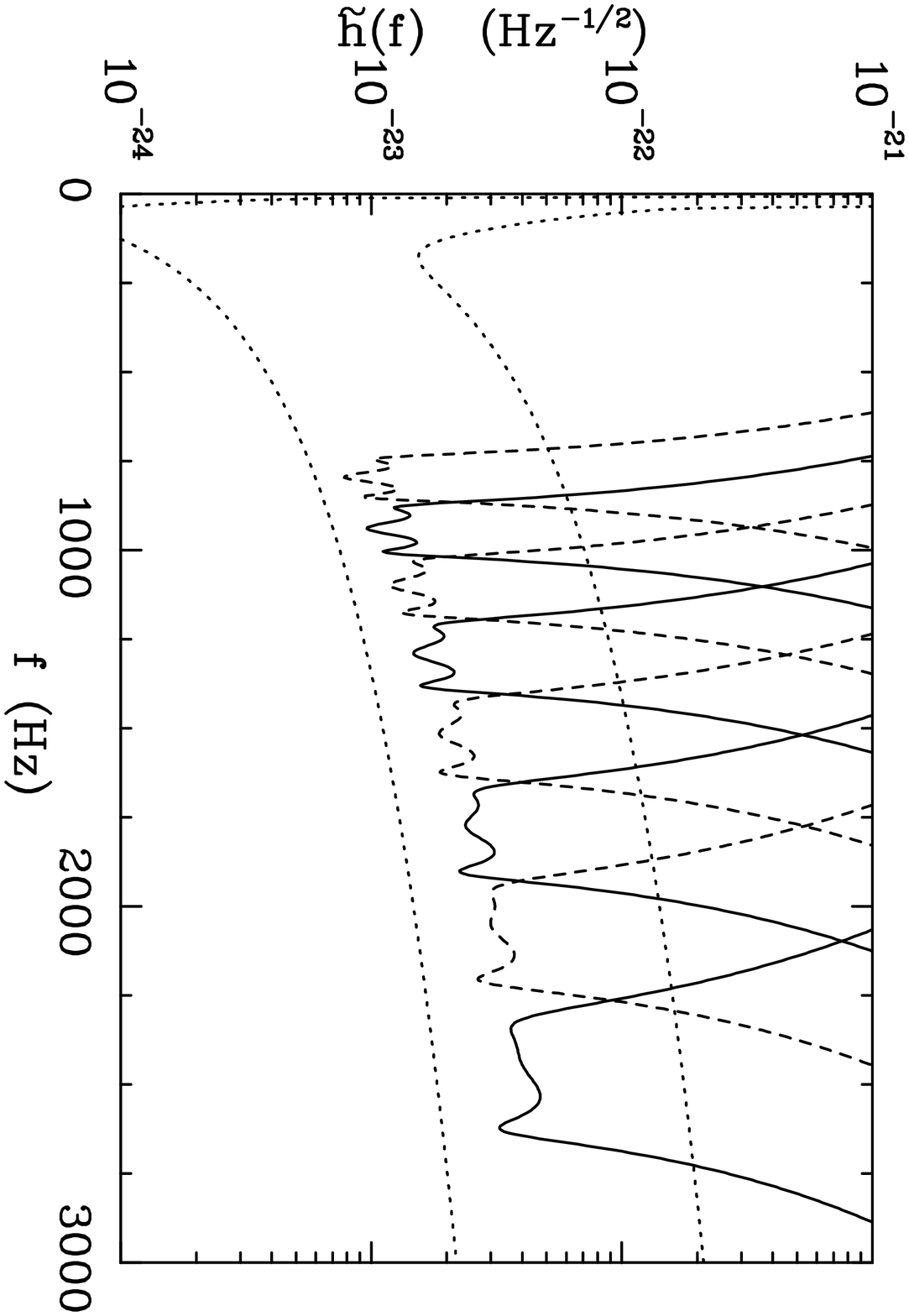}}
\end{center}
\caption{The strain spectrum of the eight spherical antennas in
the lowest quadrupole mode, shown with solid and dashed lines.  The
different line styles have no significance other than to differentiate
the separate strain spectra.  The upper dotted line shows the strain
spectrum for the first LIGO interferometer and the lower dotted line
shows the strain spectrum for the advanced LIGO interferometer, for
reference.  The spherical antennas, each with a sensitivty about
3 times the standard quantum limit, are more sensitive than the first 
LIGO interferometers in a bandwidth of 
about 100~Hz to 300~Hz each and together span a total bandwidth from 750~Hz 
to 2700~Hz. In this band, the spherical antennas are a little less 
sensitive than the advanced LIGO interferometers.
}
\label{ground}
\end{figure}

\begin{figure}[p] 
\begin{center}
\leavevmode
\rotate[l]{\epsfxsize=3.5in \epsfbox{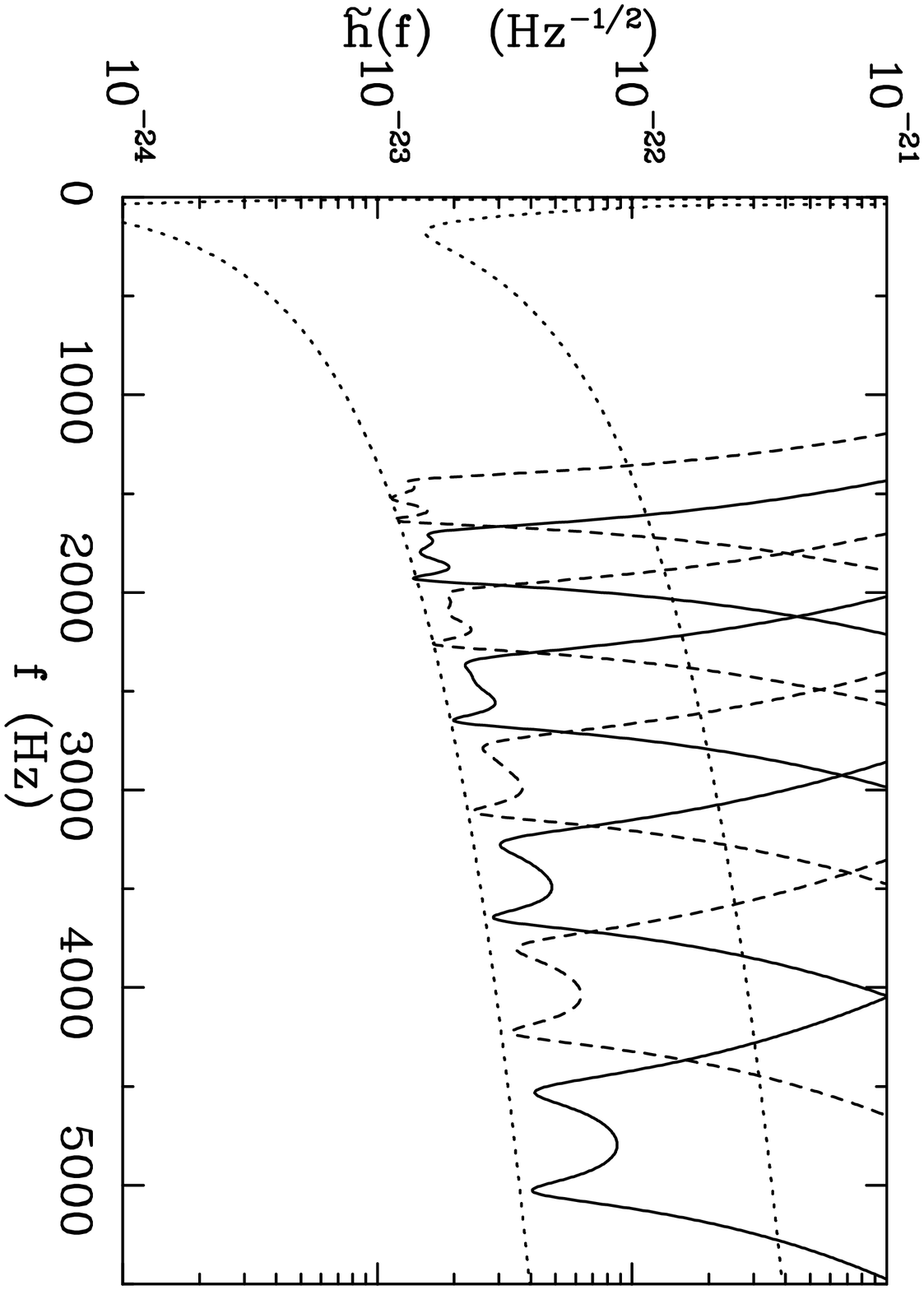}}
\end{center}
\caption{The strain spectrum of the eight spherical antennas in
the first excited quadrupole mode, shown with solid and dashed lines.  The
different line styles have no significance other than to differentiate
the separate strain spectra.  The upper dotted line shows the strain
spectrum for the first LIGO interferometer and the lower dotted line
shows the strain spectrum for the advanced LIGO interferometer, for
reference.  The spherical antennas, each with a sensitivity
about 3 times the standard quantum limit, are more sensitive than 
the first LIGO interferometers in a bandwidth of
about 200~Hz to 600~Hz each and together span a total bandwidth from 
1350~Hz to 5100~Hz.  In this band, the spherical antennas are about equal to the 
sensitivity of the advanced LIGO interferometers.
}
\label{excite}
\end{figure}

Figures~\ref{ground} and \ref{excite} show the sensitivities of the 
eight spheres in the ground
state and excited state, respectively. 
These figures also show the sensitivity of the first LIGO and advanced
LIGO interferometers for
comparison.  The data graphed is the {\em strain spectrum} of the 
detectors, $\tilde{h}(\omega)$, defined as
\begin{equation}
\tilde{h}(\omega) = \sqrt{\frac{|h(\omega)|^{2}}{\sigma(\omega)}}.
\end{equation}
The strain spectrum is a measure of what frequency distribution an 
incoming gravitational wave would have to have
in order to produce an output in a noiseless detector that 
mimics the output of the real 
detector's noise.  It is a useful way to compare detectors because it 
is independent of source 
waveform and thus is solely a characteristic of the antenna.  
The energy SNR per unit bandwidth, $\sigma(\omega)$, is related to the 
quantities $h_{\text{c}}$ and $h_{\text{rms}}$ used by LIGO and defined
in Ref. \cite{300yrs} by
\begin{equation}
\sigma(\omega) = \frac{1}{f} 
                 \left(\frac{h_{\text{c}}}{h_{\text{rms}}}\right)^{2}.
\end{equation}
Using these quantities, a value for the SNR of 
LIGO is often estimated as the maximum value of 
$(h_{\text{c}}/(\sqrt{5} h_{\text{rms}}))^{2}$, which gives a rough approximation 
to the integral in Eq. (\ref{snr}).   The factor of $\sqrt{5}$ is necessary to
convert to ``random direction and polarization'' \cite{300yrs}.

These figures show that the spherical
resonant--mass antennas have a sensitivity intermediate between 
the first and advanced LIGO interferometers within a 
fractional bandwidth of about 10\% each. The
collection of all the TIGAs, or the ``xylophone'', is a more 
sensitive detector than the first LIGO interferometers from 750~Hz 
to 2700~Hz in the lowest mode and from 1350~Hz to 5100~Hz in 
the first excited mode.  In these frequency regions LIGO's 
sensitivity is constrained by photon shot noise in the lasers 
\cite{LIGO}.

\section{Signals and Results}
\label{results}

In order to integrate $\sigma(\omega)$ and obtain the SNR for the 
spherical antenna, numerical values for a gravitational waveform 
from an astronomical event are needed.  For inspiraling and coalescing
binary neutron stars, we used the waveform published by Zhuge, 
Centrella, and McMillan \cite{Zhuge}.  We Fourier transformed
the time--domain data using the convention 
$h(\omega) = \int^{\infty}_{-\infty} 
 e^{- j \omega t} h(t) dt$.  Although the calculation by Zhuge {\em et
al.} is among the most complete available, it is still only a first
survey of how binary neutron stars may behave.  In particular, it models
gravity with a purely Newtonian formula.  The inclusion of 
general--relativistic corrections may significantly change the orbit
and lead to differences in the waveform.  Our results derived using
this waveform should be seen in this light.

\begin{figure}[p] 
\begin{center}
\vspace{0.35in}
\leavevmode
\rotate[l]{\epsfxsize=3.5in \epsfbox{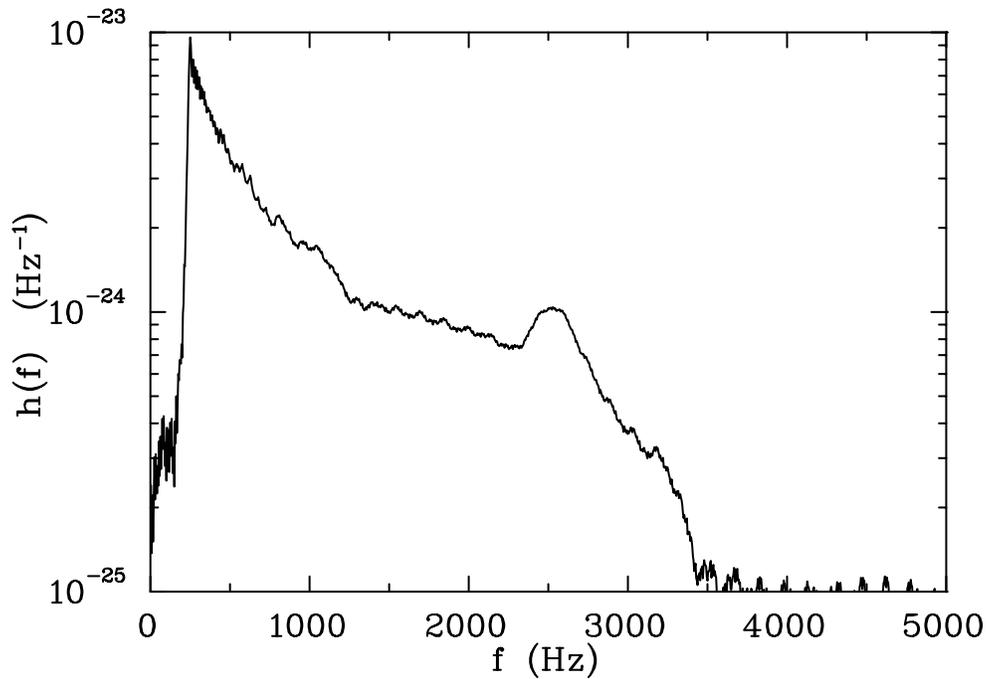}}
\end{center}
\caption{The frequency--domain gravitational waveform averaged over source
orientation from the inspiral 
and coalescence of two neutron stars with mass 
1.4 M$_{\odot}$ and radius 10~km at 15~Mpc from the antenna.  The sharp 
cut--off at 300~Hz is due to
the finite extent of the time--domain data.  The spectrum from 300~Hz 
to about 1000~Hz is due mainly to the 
inspiral phase.  The frequency $f_{\text{dyn}} = 1566$ is the dynamical 
instability frequency.  The peak at
$f \approx 2500$~Hz is associated with the transient barlike structure 
that forms immediately 
following the onset of coalescence \protect\cite{Zhuge}.}
\label{fftns}
\end{figure}

The frequency--domain waveform for the inspiral and coalescence phase 
of the binary neutron star
evolution is shown in Fig.~\ref{fftns}.  Zhuge {\em et al.} generated 
the waveform using a three--dimensional 
numerical simulation which models the neutron stars as nonrotating 
polytropes.  The neutron stars were chosen to
have equal masses of 1.4 M$_{\odot}$ each, since all known cases of 
neutron stars in binary systems
have this mass \cite{Phinney}.  Initially, the distance between the 
stars was chosen to be much larger
than the diameter of individual stars, so tidal gravitational effects 
are negligible.  Thus, the stars
are originally spherical, with a radius of 10~km.  The initial orbit 
was chosen to be nearly circular and it evolves due to 
Newtonian gravity with a frictional term added to simulate the energy 
loss to gravitational wave emission.  When the stars spiral together, 
tidal distortions in each star's shape grow larger and the evolution 
approaches coalescence.

Once the separation between the stars is comparable to the neutron star's 
radius, hydrodynamic effects 
become important and an approximation of the nuclear equation of state 
is required.  Zhuge {\em et al.} used
\begin{equation} 
P = K \rho^{1+1/n}
\label{poly}
\end{equation}
as the equation of state, where $P$ is pressure, $\rho$ is density, 
$K$ is a
 constant that measures
the specific entropy of the nuclear matter and $n$ is the polytropic 
index.  
A value of $n=1$ was 
used for the waveform we analyzed.  Smooth particle hydrodynamics (SPH) 
is then used to model the
coalescence phase once the equation of state is specified.

The gravitational waveform was calculated from the complete orbit of 
the binary 
neutron
star system using the quadrupole approximation by Zhuge {\em et al.}  
This 
approximation ignores contributions from mass moments 
higher than the quadrupole, but is valid for nearly Newtonian sources 
\cite{MTW}.  
In the 
transverse--traceless (TT) gauge, the gravitational wave amplitude is
\begin{equation}
h_{ij}^{TT} = \frac{G}{c^{4}} \frac{2}{r} \ddot{I}_{ij}^{TT},
\label{htt}
\end{equation}
where $\ddot{I}$ is the second time derivative of the reduced 
quadrupole mass 
moment of the source.  
The amplitude of the ``plus'' and
``cross'' polarizations of the gravitational wave, expressed in 
spherical 
coordinates, are  
\begin{eqnarray}
h_{+} &=& \frac{G}{c^{4}} \frac{1}{r} ( \ddot{I}_{\theta \theta} - 
          \ddot{I}_{\phi \phi}) \\
h_{\times} &=& \frac{G}{c^{4}} \frac{2}{r} \ddot{I}_{\theta \phi}. 
\end{eqnarray}
The absolute scale of these amplitudes requires a choice for $r$, 
the distance 
from the detector
to the source.  We used $r=15$ Mpc, the approximate distance to the 
Virgo cluster of galaxies \cite{300yrs}.  This distance is estimated
by Phinney \cite{Phinney} to be below the most optimistic value to
get 3 events per year, 23 Mpc.  Scaling from this value, about 1 event
per year is predicted at 15 Mpc in this optimistic limit.

The waveform from the Newtonian inspiral with friction was then
meshed onto the waveform from SPH by Zhuge {\em et al.} to get 
a complete waveform for the whole binary neutron star 
evolution.  Since the orientation angles of the binary system are not 
known {\em a priori}, and in
fact are values that the spherical antenna can determine experimentally 
\cite{Fafone}, we 
averaged the waveform over these unknown angles.  This averaging is 
done 
so that the energy per 
frequency, $dE/df$, radiated by the binary system is held constant.  
Thus, in Eq.~(\ref{energy}),
\begin{equation}
|h(\omega)|^{2} = \langle |h_{+}(\omega)|^{2} + 
                  |h_{\times}(\omega)|^{2} 
                  \rangle,
\label{h}
\end{equation}
where $\langle \cdots \rangle$ denotes an average over all source 
angles.  
It is this waveform that is 
shown in Fig.~\ref{fftns} and was used as input in Eq.~(\ref{sigma}).

Once a numerical expression for the waveform $h(\omega)$ was made 
available to us, it was possible to obtain SNRs
for the eight spheres with the diameters shown in Table~\ref{bns0}.  
To do this, the integral in 
Eq.~(\ref{snr}) must be evaluated.
Performing this integration with the $h(\omega)$ from Eq.~(\ref{h}) 
gave the SNRs in the column marked ``Total'' in Table~\ref{bns0}.  
The row marked 
``Xylophone'' is what an array of all eight TIGAs acting together could 
accomplish and is the sum
of each SNR in the rows above.  The row marked ``First LIGO Interferometers'' 
is for
comparison with the first LIGO interferometers and was calculated by using 
the same waveform integrated with 
the strain spectrum published for LIGO \cite{LIGO}.  Since the waveform 
is of finite extent in time, 
the frequency domain data is not accurate below 300 Hz. In order to get 
a reasonable value for the 
SNR of LIGO, it was necessary to extrapolate the data below this cut 
off and into LIGO's sensitive 
region.  We did this with the analytical waveform in Eq.~(44) of 
Ref.~\cite{300yrs}.  Note that this equation as published has a factor
of 2 error.  The value $\pi/12$ should be instead 
$\pi/6$ \cite{thornemail}.  We used the corrected version for
the extrapolation.  It was also necessary to divide the LIGO SNR by a
factor of 5 to represent a wave with random direction and polarization
\cite{300yrs}.
This correction is unnecessary for the spherical antennas because they
are equally sensitive in all directions.

To determine how effective the TIGAs can be in observing the 
coalescence
phase of the binary neutron star evolution, we separated the 
waveform into two
parts.  The inspiral occurs from $t=0$~s to $t=0.234$~s, and the 
coalescence occurs from 
$t=0.234$~s to $t=0.241$~s.  This division of time was chosen so that 
the instantaneous
frequency at $t=0.234$~s coincides with  $f_{\text{dyn}}$, the dynamical 
instability 
frequency identified by Zhuge {\em et al.} as the frequency where the
neutron stars cease to act as point masses. The separate time--domain 
waveforms were then multiplied 
by a Hahn windowing function \cite{window} before Fourier transforming, 
to ensure that the division 
was smooth and no spurious high--frequency signals 
were artificially created. The SNRs obtained from each of these
separate waveforms are shown in the columns marked ``Inspiral'' and 
``Coalescence'', respectively, in Table~\ref{bns0}.

With the results, we can calculate the energy sensitivity $E_{s}$
achieved by the spheres from Eq. (\ref{es}).  The energy deposited
in the 3.25 m sphere in the lowest mode is $1.79 \times 10^{-29}$
J, from Eq. (\ref{energy}).  Thus, with a SNR  of 11.3,
\begin{eqnarray}
E_{s} & = & \frac{1.79 \times 10^{-29} \text{J}}{11.3} \nonumber \\
      & = & 3.0 \hbar \omega_{0},
\end{eqnarray}
in good agreement with the approximate calculation in
Eq. (\ref{alpha}).

We also calculated SNRs for the spheres instrumented with
resonant--mass transducers tuned to the
first excited mode of the antenna.  The same physical parameters were 
used to model the spheres and
the same waveforms used as signals.  This data is shown in 
Table~\ref{bns1}.

Binary neutron star events are the best understood signals for
gravitational wave detection, but because of the high operating
frequency of spherical antennas,
other astronomical sources may be important for the spheres.  
One possible high--frequency
signal is from the 
dynamical bar--mode instability of a rapidly rotating star.  This
event may be detectable by spherical 
antennas provided the star is compact enough.  The stellar radius
that this event occurs at is uncertain. This 
instability has been investigated
by Smith, Houser, and Centrella \cite{Houser} and it is their
numerical waveform data that we used.  

\begin{figure}[p] 
\begin{center}
\rotate[l]{\epsfxsize=3.5in \epsfbox{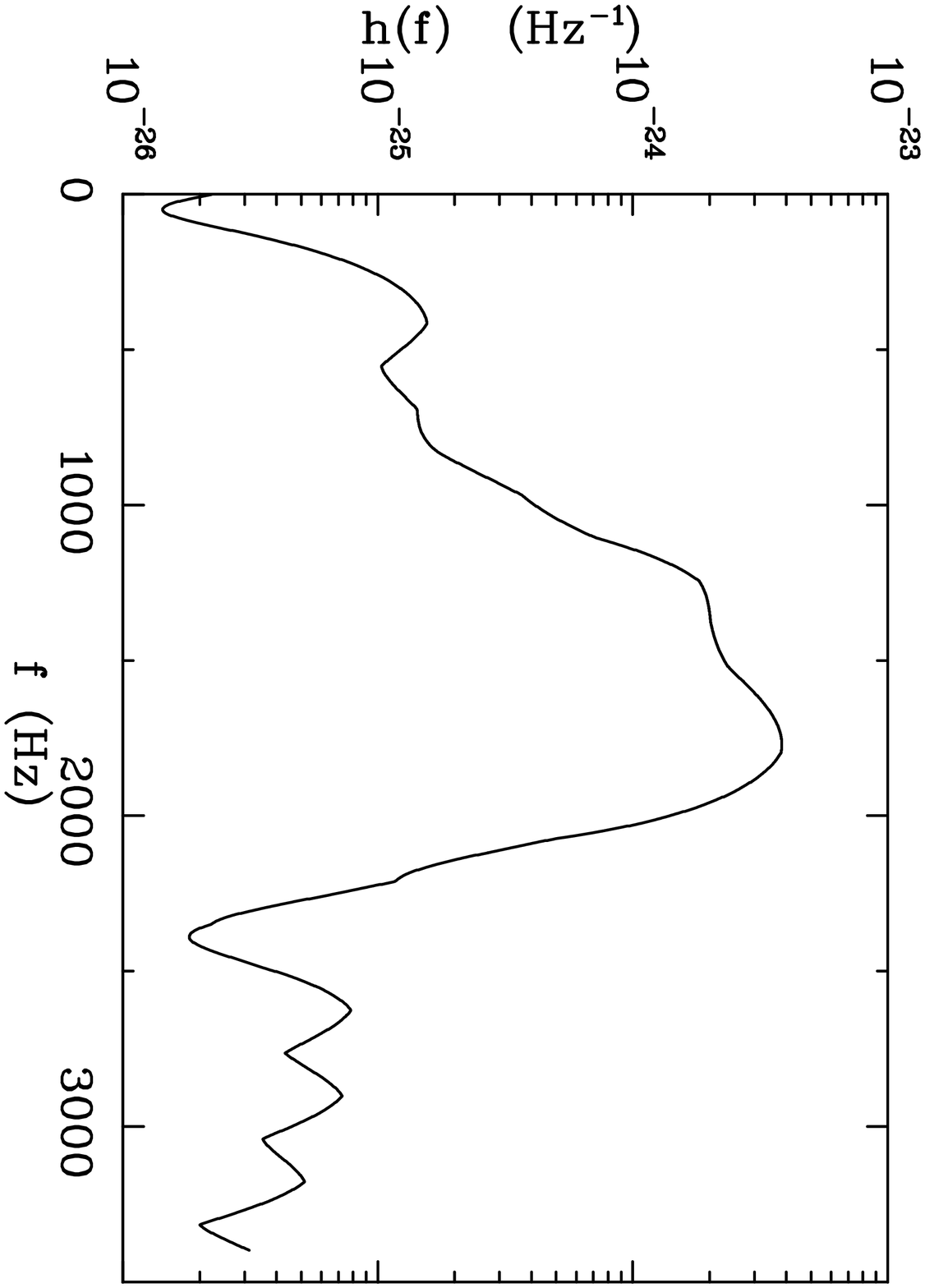}}
\end{center}
\caption{The frequency--domain gravitational waveform averaged over source
orientation from the bar--mode instability of a rapidly 
rotating star of mass 1.4 M$_{\odot}$ and radius at centrifugal hangup of
20~km at a distance of 1~Mpc from the antenna.  The 
primary peak at 1765~Hz is
twice the rotational frequency of the star.}
\label{fftrrs}
\end{figure}

Figure~\ref{fftrrs} shows the spectrum for the bar--mode instability
which develops following ``centrifugal hangup'' during the core 
collapse of a massive star.  This 
gravitational waveform was generated by Smith {\em et al.} using 
three--dimensional numerical 
simulations which modeled the star as a polytrope, with equation of 
state

\begin{equation}
P = \frac{\rho \epsilon}{n}.
\end{equation}
Here $P$ is pressure, $\rho$ is density, $\epsilon$ is specific 
internal 
energy and $n$ is the
polytropic index.  A value of $n=3/2$ was used for the waveform in our 
simulation.  A total
mass of 1.4 M$_{\odot}$ was assumed, as the star is expected to end up 
as a neutron star. An
equatorial radius for the centrifugal hangup of 20 km was also assumed. 
This is the radius where the star's collapse is arrested by the 
centrifugal effects from the rotation.
A realistic value for this radius is not known, but could range
from as high as 3000 km down to as low as 10 km \cite{Houseremail}.  The
initial rotation had a ratio of rotational 
kinetic energy to gravitational
potential energy of $\tau \approx 0.30$.  Newtonian gravity was assumed 
and the gravitational radiation
produced from the dynamical bar--mode instability was calculated in the 
quadrupole approximation, as
with the coalescence waveform.  Back reaction from gravitational wave 
emission was ignored.  
The bar mode, i.e., $m=2$ mode, was used as it is expected to be the 
fastest
growing mode \cite{Houser}.  This waveform was generated for the 
{\em dynamical} bar instability,
which is driven by Newtonian hydrodynamics and gravity rather than 
the {\em secular} instability,
which is due to dissipative processes such as gravitational radiation 
reaction. The  dynamical instability develops on 
a time scale of about one rotation period while the secular
instability grows over several periods or even more slowly.  The choice 
of $\tau = 0.30$ is just above the dynamical stability limit of 0.2738
\cite{Lai} and thus is a reasonable approximation
for a star that spins up, due to collapse or accretion, and becomes 
dynamically unstable \cite{Houser}.  
The star's evolution was simulated by Smith {\em et al.} using SPH and 
from this evolution the gravitational waveform is
calculated by using Eq. (\ref{htt}).  

The choice of $r$, the distance from source to detector, is not
as simple as for the binary neutron star.  There is less observational 
evidence for stars
with bar--mode instabilities.  Such rapidly rotating stars may be formed 
from supernovae,
so the rate of supernovae might be taken as a reasonable guide to the 
rate of this gravitational wave
event.    Out to the Virgo cluster of galaxies (15 Mpc), the supernovae 
rate is estimated at a few per year 
\cite{HouserRef}. We took the source
distance to be $r=1$~Mpc, which has an estimated event rate of 1 every 
10 years \cite{HouserRef}. Once the magnitude of each 
polarization state was evaluated, 
the same average over angles as in Eq. (\ref{h}) was performed to give 
the waveform shown in Fig.~\ref{fftrrs}.
The SNRs were calculated by using the method described in 
Section~\ref{method}.  
These values
are shown in Table~\ref{rrs} for both the ground state and the first 
excited 
quadrupole modes.  The
1.45~m, 1.25 m, and 1.05~m diameter spheres do not have data listed for 
the excited mode because the frequency--domain waveform cuts off at 
3500~Hz, which is below the resonance 
frequencies of these spheres.  This is because
the granularity of the time domain data provided was too great for 
frequencies above 3500~Hz.  However,
we believe it is safe to assume that the frequency--domain data would 
remain below $h(\omega) = 
10^{-25}$~Hz$^{-1}$ and thus the SNRs for these two spheres would be 
negligible.

The secular instability can develop
for $\tau$ values greater than 0.1375 \cite{Lai}.  After the dynamical
instability has run its course, Houser {\em et al.} find the system has
evolved to a nearly axisymmetric state with a core having $\tau=0.26$,
which is above the secular instability limit.  Lai and Shapiro \cite{Lai}
have investigated the secular instability which may develop in which
a Maclaurin spheroid evolves into a Dedekind ellipsoid producing a
gravitational wave signal sweeping in frequency from possibly near
1~kHz down to zero.  A different type
of secular evolution would apply if the calculations of Durisen {\em et
al.} \cite{Durisen} or Williams and Tohline \cite{Tohline} correctly
predict the end point of the dynamical instability as a bar surrounded 
by a ring, rather than the spheroid found by Houser {\em et al.}  If a 
rapidly spinning bar is produced, the secular evolution changes the bar
from a Jacobi--like ellipsoid into a Maclaurin spheroid with a 
gravitational wave signal from 500~Hz to as high as 3~kHz. 

Lai and Shapiro give analytical waveforms for these two different
secular instabilities.  For the same
$1.4 M_{\odot}$ star at 1~Mpc with a radius of 10~km averaged over
all source and detector angles, the Dedekind
waveform would give an energy SNR of $2000$ in the 3.25~m diameter sphere and
$1200$ for the first LIGO interferometers.  Such a
strong signal would be detectable even at the Virgo cluster distance.  
However, it seems likely that the starting frequency for the Dedekind
evolution will be well below the spherical detector's band unless the star
is initially spinning just below the dynamical instability limit and is
nearly incompressible.
In order for the starting frequency of this event to exceed the 
795~Hz frequency of the 3.25~m sphere, Lai and Shapiro's Fig.~(5) indicates 
one needs a polytropic index $n < 0.7$ for $\tau=0.26$, while even $n=0$ does 
not suffice for $\tau \leq 0.24$.  In contrast, the frequency span of the
signal from Jacobi--like evolution seems to have a higher possibility of
overlap with the band covered by the
spheres; however, the chief uncertainty for that signal is whether or not
the dynamical instability produces a spinning bar rather than a spheroid.  
{\em If} Jacobi-like evolution occurs and its starting frequency is below
800~Hz, then the signal computed by Lai and
Shapiro is a very strong source for a big sphere:  we calculate an energy 
SNR of $4000$  for a 3.25~m diameter sphere and an energy SNR of $240$ 
for the first LIGO interferometers from a waveform averaged over all angles
and a source distance of 1~Mpc.

\section{Conclusion}
\label{con}
The results in Table~\ref{bns0} for the spherical antenna tuned to the 
lowest quadrupole 
frequency interacting with gravitational radiation from binary neutron 
stars shows
that spherical antennas operate at a level that is complementary with the
first LIGO interferometers.  The largest sphere
obtains an energy SNR of 11.3 at a distance of 15 Mpc.   If a
SNR of 10 in this detector is a sufficient threshold for a three--way 
coincidence experiment, then a source with angle averaged strength 
could be detected out to a distance of 15.9 Mpc.  
With the most optimistic estimate of the coalescence rate, 3 events 
per year out to 23 Mpc \cite{Phinney},  the expected rate of detection 
$r_{d}$ is
\begin{eqnarray}
r_{d} & = & 1.15 \times 3 \: \text{yr}^{-1} \left( \frac{15.9\,
\text{Mpc}}{23 \, \text{Mpc}} \right)^{3} \\
   & = & 1.15 \: \text{yr}^{-1}.
\end{eqnarray}
The factor of 1.15 
is due to statistical preference for angles that give high SNRs 
(see \cite{300yrs} for details).   For detection of 
gravitational radiation from binary neutron stars at a distance of
15 Mpc, a 3.25~m diameter aluminum 
sphere near the standard quantum limit will be sufficient.   The upper
limit on the event rate at this distance is about 1 coalescence per
year \cite{Phinney}.

Table~\ref{bns0} also shows that a large sphere instrumented at the 
lowest quadrupole frequency
does not hold out much hope of seeing the details of binary neutron
star coalescence.  Even the 1.05~m
diameter sphere, whose size was chosen so that the lowest quadrupole
mode was at the maximum of the
coalescence wave spectrum, does not manage to reach a SNR of 1.  As
the frequency of the sphere goes
up, the radius, and with it its mass, goes down.  At frequencies where 
the waveform from Zhuge {\em
et al.} is strong, the energy cross section of the sphere is too small 
to detect much.  This raises
the question of the reliability of the numerical waveform, especially 
of the frequency $f_{\text{peak}}$
associated with the barlike transient.  According to Centrella 
\cite{Centrella}, the qualitative 
shape of the waveform is fairly reliable, but the exact position of 
this peak and other structures may 
change as numerical relativity techniques improve.  
If $f_{\text{peak}}$ were to be found at a lower 
frequency, closer to the lowest quadrupole mode of one of the 
larger spheres, the prospect for a SNR 
greater than 1 for the coalescence phase might improve.

The data in Table~\ref{bns1} for spheres sensitive at the first 
excited quadrupole 
mode to inspiraling
and coalescing binary neutron stars appears a little more promising 
for the 
detection of coalescence.  The
largest sphere still has the highest overall SNR, but it is reduced 
from its value 
in the ground state.
The 2.00~m diameter sphere, with an excited quadrupole frequency of 
2483~Hz, is 
now the sphere  
tuned to $f_{\text{peak}}$.  It does not quite manage a SNR of 1 
either, but it 
does have a higher SNR
for the coalescence than for the inspiral phase, as do the 1.70~m 
and 1.45~m spheres.

The data in Tables~\ref{bns0} and \ref{bns1}, taken together, suggest 
that a 
xylophone of spheres
acting collectively and in collaboration with LIGO may be the best 
approach to detection of binary neutron 
star coalescence.  The smaller
spheres do not contribute much to the xylophone SNR, so four spheres, 
from 3.25~m to 2.00~m in diameter, would
be enough to give most of the xylophone benefits.  If these four
spheres could be instrumented at both
the ground and the first excited quadrupole frequencies, a fairly 
wide spectrum, from 750~Hz to 2700~Hz continuously, could be monitored.  
The spheres could rely on LIGO
to provide a high SNR detection of the inspiral.  Once the event
is discovered, the spheres would have the less strenuous task of 
identifying and analyzing the high frequency coalescence.  Thus,
even the modest SNRs found in our work may still prove useful in 
gathering astrophysical data on coalescing binary neutron stars.

The data for the rapidly rotating star shown in Table~\ref{rrs} is 
more encouraging.  For the ground
state modes, the 1.45~m sphere has its frequency near the peak of 
the spectrum and obtains a strong SNR
of almost 10.  The two spheres with quadrupole frequencies below 
the resonance of the 1.45 m sphere 
also have strong SNRs, about 7 and 5.  Further away from the peak, 
SNRs fall off rapidly, especially on the
high--frequency end.  The SNR goes from 0.935 for a sphere diameter of 
1.25~m to well below 1 for a diameter
of 1.05~m.  Thus with only two antennas the peak signal can be easily 
found, provided their 
sensitive frequencies occur at the appropriate positions.

The first excited state data is similar to the ground state, showing 
strong SNRs when the sphere's
quadrupole frequency is at or near the 1765~Hz peak.  In the excited 
mode, however, this occurs between
the 3.25~m and the 2.75~m diameter spheres. With larger masses, these 
spheres have higher energy cross 
sections and thus obtain much higher SNRs.  A SNR of 25, from the 
2.75~m sphere in the excited state, 
represents such a strong signal that the source position on the 
sky could be located to within almost
0.13~steradian \cite{ZM}.  As with the ground state data, there is 
a sharp drop in SNR to effectively 
zero about 500~Hz above the peak frequency, making location of 
the peak frequency possible.

The rapidly rotating star waveform was generated by assuming a total mass 
of 1.4 M$_{\odot}$ and a centrifugal hangup at 20~km.  The location of 
the peak frequency, which 
is twice the rotation frequency of the
star, is very sensitive to the values of these parameters.  
It can be as low as 1~Hz for
a 1.0 M$_{\odot}$, 3000~km star up to 6000~Hz for a 2.0 M$_{\odot}$, 
10~km star \cite{Houseremail}.  Since
the appropriate values for these star parameters are not known, 
and in fact are values we could hope to 
determine from gravitational wave data, actual signals from this 
source could potentially be outside
the sensitive range of spherical antennas.  A peak frequency above 
about 2500~Hz, corresponding to a 
1.4 M$_{\odot}$, 10~km star, would be extremely difficult to detect 
outside our galaxy.  This would limit
the number of events to a few per century.
Detecting these higher frequency signals depends on the accuracy of the 
current template, especially the
secondary peak in Fig.~\ref{fftrrs} at 400~Hz.  Many details of the 
rapidly rotating star's evolution
are not well understood and this waveform may undergo substantial 
changes as the field of numerical relativity advances.

A xylophone of a 3.25~m, a 2.75~m, a 2.35~m and a 2.00~m diameter sphere 
instrumented at both the ground state and the
first excited quadrupole state working along with LIGO, as suggested for 
the inspiraling and 
coalescing binary neutron stars,
would do an effective job of searching for rapidly rotating star events.  
For favorable mass and radius 
parameters, the excited state of the 2.75~m diameter sphere would be
sensitive, with a SNR threshold of 10, out to 1.6~Mpc.  This will be 
sufficient to observe one event every decade, provided nearly all
type I$_{\text{b}}$ and type II supernovae are sources of this event
\cite{HouserRef}.  This xylophone 
would also determine the star's rotation frequency, as a large SNR
would be seen in the more massive 
sphere and effectively nothing would be seen in the smaller.  This 
would locate the peak frequency, 
and hence the rotation frequency, to within a few hundred hertz. 
Negative results from such a 
xylophone would restrict the parameter space, providing data 
about neutron star development and equation of state.  The
xylophone covers a frequency range corresponding to hangup radii between
45~km down to 15~km.  The broadband interferometers have difficulty 
detecting this dynamical instability, but do have a high SNR for the 
secular instability.  Since these two gravitational wave events may
occur in succession, a collaborative effort between spheres and 
interferometers would prove effective as with the binary neutron stars.

In this paper, we have compared broadband interferometers with 
resonant--mass antennas for detection of high frequency gravitational 
radiation.  Another possible technique for detecting high frequency 
events involves dual--recycled interferometers 
\cite{krolak}.  This approach allows laser interferometers to become
much more sensitive within a narrow bandwidth at the expense
of sensitivity elsewhere.  Krolak, Lobo, and Meers \cite{krolak} have 
looked at SNRs for inspiraling neutron stars
interacting with recycled interferometers using a simplified strain
spectrum and an analytical formula for the waveform.  Their results for
the unrecycled LIGO are in close agreement with ours.  Further
investigations of SNRs comparing spherical antennas with dual--recycled
interferometers is being done at Caltech \cite{thornemail}. 

In addition to the astronomical sources of gravitational radiation that 
we investigated here, there
may be other high frequency signals potentially detectable in detectors
optimized for the 1--2 kHz band.  Events that might produce high--frequency 
gravitational waves include coalescence 
of a neutron star with a black hole or a black hole with a second
black hole \cite{Thorne}, asymmetric core collapse and bounce in 
supernovae \cite{Thorne}, spinning neutron stars \cite{Zimmerman}, and
cosmic string vibrations 
\cite{Vilenkin}.  Especially promising may be the black hole 
coalescences and spinning neutron 
stars.  Excitation of the high--frequency (f~$\approx 1200-3250$~Hz 
$(10 M_{\odot}/M)$ \cite{blackhole}) 
black hole quasi--normal modes would give a relatively strong 
signal at kilohertz frequencies.  
A gravitational--wave antenna detection of this radiation could provide 
the observational ``smoking gun'' 
to confirm the existence of black holes.  Experimental evidence of 
gravitational radiation from black 
hole coalescence would undoubtedly also provide great insight into
relativistic gravity.  Spinning 
neutron stars are a periodic source that could radiate strongly for 
months \cite{Zimmerman}.  The frequency of the
waves would be twice the rotation period, often above a kilohertz, 
and details of the wave would 
tell much about the structure of neutron stars.  According to Thorne
\cite{Thorne}, ``the deepest searches 
for these nearly periodic waves will be performed by narrow--band 
detectors ... e.g., dual recycled 
interferometers or resonant--mass antennas.''
We call on the numerical relativity community to continue 
to develop reliable waveforms for all
possible high--frequency events.  It is only through the combined 
efforts of everyone; interferometer and
resonant--mass experimentalists, as well as numerical and analytical 
theorists, that confirmed, direct
detection of gravitational radiation will become a reality. 

\section*{Acknowledgments}
We would like to thank J. M. Centrella and  J. L. Houser for providing us 
with the numerical data for their gravitational waveforms.
We also thank J. M. Centrella, J. L. Houser, S. A. Hughes, C. W. Misner, 
J.-P. Richard, S. R. Sinha, K. S. Thorne, F. W. Wellstood, and X. Zhuge for 
helpful discussions as well 
as R. K. Burrows and P. C. Dulany for invaluable computer assistance 
and comments.  This work was supported by NSF Grant No. PHY93-12220.

\pagebreak
\setcounter{equation}{0}
\renewcommand{\theequation}{A\arabic{equation}}
\section*{APPENDIX:  Asymptotic expansion of the energy sensitivity}

Equation (\ref{es}) defines the energy sensitivity $E_{s}$ as the ratio of $E$, the
energy the signal would deposit in a bare antenna initially at rest, to the
SNR calculated by integrating $\sigma(\omega)$ given by Eq. (\ref{sigma}).  Although we
calculated $E_{s}$ directly in this fashion, useful insight into the 
dependence of $E_{s}$ on the detector parameters comes from considering the
asymptotic behavior for the $Q_{i}$ approaching infinity as indicated by 
Eq. (\ref{alpha}).

First, consider the lossless case when all Qs are infinite.  Then 
$\text{Re}(y_{22})$ is zero, and Price \cite{Price} shows that the expression
for $\sigma(\omega)$ can be rewritten exactly as

\begin{equation}
\sigma(\omega) = \frac{e_{n}(\omega)}{k_{B} T_{n}},
\label{en1}
\end{equation}
where 
\begin{equation}
e_{n}(\omega)  =  \frac{|f_{1} y_{21}|^{2}}{|1 + z_{n} y_{22}|^{2}} r_{n},
\label{en2}
\end{equation}
$T_{n}$ is the mechanical amplifier noise temperature defined by Eq. (\ref{tn}), and
\begin{equation}
z_{n} = r_{n} + j x_{n}
\end{equation}
is the mechanical amplifier's complex noise impedance, defined by 
Eqs.~(\ref{rn},\ref{xn}).

As Price \cite{Price} explains, $e_{n}(\omega)$ has the following physical
interpretation:  it is the spectrum of the energy which would be dissipated in
$r_{n}$ if the signal were applied when the amplifier had been replaced by its
input impedance plus an additional mechanical impedance equal to $z_{n}$.  If 
$f_{1}(\omega)$ varies little over the band where $e_{n}(\omega)$ is large,
then the signal is approximately an impulsive force which deposits the energy

\begin{equation}
E = \frac{|f_{1}(\omega_0)|^{2}}{2 m_{\text{eff}}}
\end{equation}
initially all in antenna motion.  Since the hypothetical detector with $z_{n}$
inserted is assumed lossless except for $r_{n}$, eventually 100\% of the energy
$E$ ends up being dissipated in $r_{n}$.  Thus from this physical argument we
have 

\begin{equation}
\int_{-\infty}^{\infty}\frac{d \omega}{2 \pi} e_{n}(\omega)
 \approx \frac{|f_{1}(\omega_0)|^{2}}{2 m_{\text{eff}}},
\label{enint}
\end{equation}
independent of the values of $r_{n}$ and the detector masses and springs. 
Therefore, in the lossless case, the SNR is 

\begin{eqnarray}
\text{SNR} & = & \int_{-\infty}^{\infty}\frac{d \omega}{2 \pi} \sigma(\omega) \\
    & = & \int_{-\infty}^{\infty} 
                 \frac{d \omega}{2 \pi} \frac{e_{n}(\omega)}{k_{B} T_{n}} \\
    & \approx & \frac{E}{k_{B} T_{n}},
\end{eqnarray}
provided $f_{1}$ and $T_{n}$ vary little over the frequencies where $e_{n}$ is
significant.  Hence, for an impulsive force, $E_{s} = k_{B} T_{n}$ in the
lossless case independent of the other detector parameters.

Next, consider the case where all Qs are infinite except that for the antenna.
Then the antenna dissipation produces a thermal force noise acting on 
$m_{\text{eff}}$ with spectral density $2 k_{B} T m_{\text{eff}} 
\omega_{0}/Q_{\text{eff}}$.
Therefore, Eq. (\ref{sigma}) becomes

\begin{equation}
\sigma(\omega) = \frac{|f_{1} y_{21}|^{2}}{S_{u} + S_{f} |y_{22}|^{2} +
                    2 \text{Re}(y_{22} S_{fu}) + 
		    2 k_{B} T \frac{m_{\text{eff}} \omega_{0}}{Q_{\text{eff}}} 
                    |y_{21}|^{2} }.
\label{newsig}
\end{equation}
For convenience, we define the energy sensitivity expressed in temperature 
units to be $T_{p}$, the detector's pulse-detection noise temperature:

\begin{equation}
k_{B} T_{p} = E_{s}.
\end{equation}
Motivated by Eq. (26) in Price \cite{Price}, we use Eqs.~(\ref{en1},\ref{en2}) 
to rewrite Eq.~(\ref{newsig}) in terms of $e_{n}(\omega)$, and then expand the 
integral for SNR to first order in $Q_{\text{eff}}^{-1}$ and use Eq.~(\ref{enint}):

\begin{eqnarray}
\text{SNR} & = & \frac{E}{k_{B} T_{p}} \\
           & = & \int_{-\infty}^{\infty} 
                 \frac{d \omega}{2 \pi} \frac{e_{n}(\omega)}{k_{B} T_{n}}
                 \frac{1}{1 + \frac{T \omega_{0}}{T_{n} Q_{\text{eff}}} 
                 \frac{e_{n}(\omega)}{E}} \\
           & \approx & \frac{E}{k_{B} T_{n}} \left( 1 - 
                       \frac{T \omega_{0}}{T_{n} Q_{\text{eff}} 2 \delta f_{0}} 
                                              \right),                    
\end{eqnarray}
where
\begin{equation}
2 \delta f_{0} = \frac{\left[ \int_{-\infty}^{\infty} \frac{d \omega}{2 \pi}
                              e_{n}(\omega) \right]^{2}}
                      {\int_{-\infty}^{\infty} \frac{d \omega}{2 \pi}
                             \left[e_{n}(\omega)\right]^{2}}
\end{equation}
defines the fractional bandwidth $\delta$ of the lossless detector in a
way which arises naturally in the SNR expansion, and which agrees with the 
intuitive definition of bandwidth if $e_{n}(\omega)$ is box-shaped.

Thus when all other Qs are infinite, the asymptotic form of $E_{s}$ for large
$Q_{\text{eff}}$ is described by
\begin{equation}
\frac{1}{T_{p}} \approx \frac{1}{T_{n}} \left( 1 - 
                        \frac{T \omega_{0}}{T_{n} Q_{\text{eff}} 2 \delta f_{0}}
                                        \right)
\end{equation}
\begin{equation}
T_{p} \approx T_{n} + \frac{T}{Q_{\text{eff}} \delta} \pi 
              \; \; \; \text{as} \; Q_{\text{eff}} \rightarrow \infty.
\label{assym}
\end{equation}
When other Qs besides $Q_{\text{eff}}$ are also finite, analysis along the 
lines above is more cumbersome, but one finds
\begin{equation}
T_{p} \approx T_{n} + 
	\frac{T}{\delta} \sum_{i=1}^{N} \frac{1}{\alpha_{i} Q_{i}},
\end{equation}
where the constants $\alpha_{i}$ can be determined numerically.  Comparison
with Eq.~(\ref{assym}) shows that $\alpha_{1} = 1/\pi$ always.  Thus, a higher
fractional bandwidth $\delta$ reduces the effect of finite Qs on $T_{p}$. 

Price \cite{Price} argues that a clever choice of the masses
and springs to maximize $\delta$ for a given value of $r_{n}$ is the optimal
design strategy, and that a design in which $e_{n}(\omega)$ is made ``optimally
flat'' is nearly optimal in this sense.  However, in this paper, we have chosen 
the simpler design originally proposed by Richard \cite{Richard} which has 
constant successive mass ratios since we find both designs give similar values
for $T_{p}$.

The values of 
$\alpha_{i}$ for $i \geq 2$ depend on the masses and springs chosen for the
detector design.  The values for $\alpha_{1}, \alpha_{2},$ and $\alpha_{3}$ 
that we calculated and used are shown in Table~\ref{alphatable}.  These values
are for the case of a constant mass ratio between stages of 100.  This ratio 
was chosen as a 
convenient value that gives a final transducer mass slightly higher than the
mass that gives the lowest $T_{p}$.  A higher mass is more resistant to 
uncertainties in noise
resistance and thus is preferred experimentally to a lower than optimal mass.
The design rule suggested by Richard \cite{Richard} for choosing a constant 
mass ratio given an $r_{n}$ and $m_{\text{eff}}$ can be written as 
\begin{equation}
\left( \frac{m_{i+1}}{m_{i}} \right)^{\frac{2 N - 1}{2}} = \zeta
                   \frac{r_{n}}{m_{\text{eff}} \, \omega_{0}},
\end{equation}
where $N$ is the number of modes of the system and $\zeta$ is
2.  The design we chose corresponds to $N = 3$ and $\zeta = 2.61$; the
$\alpha_{i}$ in Table~\ref{alphatable} apply for any such design.

\begin{table}[p]
\caption{Energy signal-to-noise ratios for binary neutron star evolution in
the lowest $\ell = 2$ mode of the sphere.  The distance to the neutron
stars is taken to be 15 Mpc and each neutron star has a mass of 
1.4~M$_{\odot}$ and a radius of 10~km.  The waveforms have been averaged
over source and detector angles.  The individual spheres have a 
sensitivity about 3 times the standard quantum limit.  The row labeled 
``Xylophone'' is
obtained by summing the signal--to--noise ratios for each sphere.} 
\label{bns0}
\begin{tabular}{crlll}
Diameter& \multicolumn{1}{c}{Frequency}& \multicolumn{1}{c}{Coalescence} 
&\multicolumn{1}{c}{Inspiral}&\multicolumn{1}{c}{Total} \\
\hline
 3.25 & 795 Hz &\hspace{1em} 0.0113 &\hspace{-0.5em}10.6 &
\hspace{-0.5em}11.3 \\
 2.75 & 940 Hz &\hspace{1em} 0.00985& 5.79 & 6.28 \\
 2.35 & 1100 Hz &\hspace{1em} 0.0146 & 3.43 & 3.88\\
 2.00 & 1292 Hz &\hspace{1em} 0.00948 & 1.40 & 1.64 \\
 1.70 & 1520 Hz &\hspace{1em} 0.00853 & 0.907 & 1.09 \\
 1.45 & 1782 Hz &\hspace{1em} 0.0104 & 0.558 & 0.719 \\
 1.25 & 2096 Hz &\hspace{1em} 0.0197 & 0.285 & 0.449 \\
 1.05 & 2461 Hz &\hspace{1em} 0.126 & 0.0886 & 0.407 \\
\hline
\multicolumn{2}{l}{Xylophone} &\hspace{1em} 0.210 &\hspace{-0.5em}22.8 
&\hspace{-0.5em}25.6 \\
\multicolumn{2}{l}{First LIGO Interferometers} &\hspace{1em} 0.00406
&\hspace{-0.5em}58.2 &\hspace{-0.5em}58.8 \\
\end{tabular}
\end{table}

\begin{table}[p]
\caption{Energy signal-to-noise ratios for 
binary neutron star evolution in the
 first excited $\ell = 2$ mode of the sphere.  The waveforms have been 
averaged over source and detector angles.  
The distance to the neutron
stars is taken to be 15 Mpc and each neutron star has a mass of 1.4~M$_{\odot}$
and a radius of 10 km. The individual spheres have a sensitivity about 3
times the standard quantum limit.  The row labeled ``Xylophone'' is
obtained by summing the signal--to--noise ratios for each sphere.} \label{bns1}
\begin{tabular}{crlll}
Diameter& \multicolumn{1}{c}{Frequency}& \multicolumn{1}{c}{Coalescence} 
&\multicolumn{1}{c}{Inspiral}&\multicolumn{1}{c}{Total} \\
\hline
 3.25 & 1528 Hz &\hspace{1em} 0.0235 & 2.43 & 2.93 \\
 2.75 & 1806 Hz &\hspace{1em} 0.0289 & 1.45 & 1.88\\
 2.35 & 2113 Hz &\hspace{1em} 0.0593  & 0.679 & 1.13\\
 2.00 & 2483 Hz &\hspace{1em} 0.349 & 0.222 & 1.09 \\
 1.70 & 2921 Hz &\hspace{1em} 0.0688 & 0.0448 & 0.224 \\
 1.45 & 3425 Hz &\hspace{1em} 0.00709 & 0.00591 & 0.0252\\
 1.25 & 3973 Hz &\hspace{1em} 0.00111 & 0.00150 & 0.00513 \\
 1.05 & 4729 Hz &\hspace{1em} 0.000674 & 0.000733 & 0.00279 \\
\hline 
\multicolumn{2}{l}{Xylophone} &\hspace{1em} 0.538 & 4.83 & 7.29 \\
\multicolumn{2}{l}{First LIGO Interferometers} &\hspace{1em} 0.00406 
&\hspace{-0.5em}58.2 &\hspace{-0.5em}58.8 \\
\end{tabular}
\end{table}

\begin{table}[p]
\caption{Energy signal-to-noise ratios for the 
rapidly rotating star waveform 
for $\ell = 2$ modes of the sphere.  The waveforms have been 
averaged over source and detector angles.  
The dashes for the 
excited mode of the 1.45 m, 1.25 m and
1.05 m spheres represent the fact that the signal data cuts off below 
the resonance frequency of these modes.  The distance to the
star is taken to be 1 Mpc, the mass of the star is taken to be 1.4~M$_{\odot}$,
and the radius of centrifugal hangup is taken to be 20~km. The individual spheres
have a sensitivity about 3
times the standard quantum limit.  The row labeled ``Xylophone'' is
obtained by summing the signal--to--noise ratios for each sphere.} \label{rrs}
\begin{tabular}{crlrl}
Diameter& \multicolumn{2}{c}{Ground State}
& \multicolumn{2}{c}{Excited State} \\
        & \multicolumn{1}{c}{Frequency} & \multicolumn{1}{c}{SNR} 
& \multicolumn{1}{c}{Frequency} &
\multicolumn{1}{c}{SNR} \\
\hline
 3.25 & 795 Hz &\hspace{0.8em} 0.0661 & 1528 Hz & \hspace{0.5em}19.5 \\
 2.75 & 940 Hz &\hspace{0.8em} 0.220  & 1806 Hz & \hspace{0.5em} 25.0 \\
 2.35 & 1100 Hz &\hspace{1em}1.22   & 2113 Hz & \hspace{1em} 1.08\\
 2.00 & 1292 Hz &\hspace{1em}4.75   & 2483 Hz & \hspace{1em} 0.00434\\
 1.70 & 1520 Hz &\hspace{1em}6.95   & 2921 Hz & \hspace{1em} 0.00278\\
 1.45 & 1782 Hz &\hspace{1em}9.91   & 3425 Hz & \hspace{1em}   - \\
 1.25 & 2096 Hz &\hspace{1em}0.935   & 3973 Hz & \hspace{1em}   - \\
 1.05 & 2461 Hz &\hspace{1em}0.00168   & 4729 Hz & \hspace{1em}   - \\
\hline
\multicolumn{2}{l}{Xylophone} &\hspace{0.5em}24.1 &  &\hspace{1em}45.6 
\\
\multicolumn{2}{l}{First LIGO Interferometers} &\hspace{1em} 0.197 &  
&\hspace{1em} 0.197 \\
\end{tabular}
\end{table}

\begin{table}[P]
\caption{Coefficients in the asymptotic expansion of
energy sensitivity.  The $\alpha_{i}$ determine the influence of the Qs on
the noise temperature.  The value $\delta$ is the fractional bandwidth of 
the detector in the lossless limit.} \label{alphatable}
\begin{tabular}{ccc}
Constant & Calculated & Theoretical \\
\hline
$\alpha_{1}$          & 0.34              & $1/\pi$ = 0.32     \\
$\alpha_{2}$          & 0.41              &  --                  \\
$\alpha_{3}$          & 0.40              &  --                  \\
$\delta    $          & 0.103             & 
                            $\sqrt{m_{\text{trans}}/m_{\text{int}}} = 0.100$
\end{tabular}
\end{table}

\vspace{1in}

\vspace{1in}

\begin{thebibliography}{99}
\bibitem{coop}An American collaboration among researchers at the 
             University of 
             Maryland, Louisiana State University, the University of 
             Rochester 
             and the University of Santa Clara has been formed to 
             design and 
             build spherical antennas.  There are also
             collaborations in Italy, the Netherlands and Brazil.
\end{thebibliography}
\end{document}